\documentclass[preprint,pra]{revtex4}

\input{tcilatex}
\newtheorem{acknowledgement}[theorem]{Acknowledgement}
\begin{document}

\preprint{}
\title[Simulation of Quantum Entanglement]{Classical Simulation of Quantum
Entanglement using Optical Transverse Modes in Multimode Waveguides}
\author{Jian Fu$^{1}$, Zhijian Si$^{1}$, Shaofang Tang$^{2}$, and Jian Deng$%
^{1}$}
\affiliation{$^{1}$State Key Lab of Modern Optical Instrumentation, Department of Optical
Engineering, Zhejiang University, Hangzhou 310027, China\\
$^{2}$College of Science, Hangzhou Teachers' College, Hangzhou 310036, China}

\begin{abstract}
\quad We discuss ``mode-entangled states'' based on the optical transverse
modes of the optical field propagating in multi-mode waveguides, which are
classical simualtion of the quantum entangled states. The simualtion is
discussed in detail, including the violation of the Bell inequality and the
correlation properties of optical pulses' group delays. The research on this
simulation may be important, for it not only provides useful insights into
fundamental features of quantum entanglement, but also yields new insights
into quantum computation and quantum communication.
\end{abstract}

\date{today}
\startpage{1}
\maketitle

\section*{Introduction}

So far, there is interest in research on classical wave analogs of the
Schrodinger wave function \cite{kov}\cite{Nienhuis}\cite{dragoman}. It is
well known that, in the paraxial approximation, the transverse modes of an
optical field obey a propagation equation which is formally identical to the
Schrodinger equation with the time replaced by the axial coordinate \cite%
{kov}. The transverse modes of the optical field propagating in a waveguide
with a parabolic refractive index profile are formally identical to quantum
harmonic oscillator wave functions. Some efforts have gone into researching
on classical wave analogs of quantum mechanics, including analogs of Fock
states and measurement of Wigner phase-space distributions for classical
optical fields which can exhibit negative regions \cite{dragoman}\cite%
{iaconis}\cite{cheng}\cite{lee1}. However, research on classical analogs has
been limited principally to measurement of first order coherence, i.e.,
single-particle states. Classical-wave analogs of high order coherence
(quantum entanglement), i.e., multiparticle states, have been seldom studied %
\cite{Lee}. The quantum entanglement, which describes nonlocal quantum
correlation between different degrees of freedom especially separated
particles, is regarded as the inherent feature of quantum theory \cite%
{chuang}. The quantum correlation has been shown in the correlation
measurement of the entangled state, and a criterion has been given by the
violation of the Bell inequality \cite{bell}. In recent research, the
quantum entanglement is considered as a key property to realize the quantum
computation \cite{lidar} and quantum teleportation \cite{bennett}, which
makes the quantum entanglement strongly attracted to researchers.

In this paper, we will propose ``mode-entangled states'' based on the
optical transverse modes of the optical field propagating in multimode
waveguides. It is well-known that the quantum entanglement is the
characteristic of the quantum theory with no classical analog. Therefore,
the ``mode-entangled states'' should be interpreted as the classical
simulation of quantum entanglement using the optical transverse modes of the
optical fields. The classical simulation will be discussed in detail,
including the violation of the Bell inequality and the correlation
properties of optical pulses' group delays. In \cite{jianfu}, a full optical
scheme to perform quantum computation is proposed, based on the optical
transverse modes in multimode waveguides. The proposed C-NOT gate has the
potential of being easily realized since it is based on optical waveguide
technology and can be constructed by using Mach-Zehnder interferometer
having semiconductor optical amplifiers (SOAs) in its arms. The SOA can
provide a very large Kerr-like nonlinearity even at relatively low light
intensities and can avoid the intensity attenuation excited by two-photon
absorption \cite{soa}. Therefore SOAs have been used extensively as
nonlinear elements in optical switching and wavelength-conversion devices %
\cite{Cotter}. But whether the scheme would be capable of implementing the
quantum computation depends on whether the C-NOT gate proposed in \cite%
{jianfu} can generate the ``mode-entangled states''. Given all these, the
research on the similarity between the ``mode-entangled states'' and the
quantum entangled states may be important, for it not only provides a useful
insight into fundamental features of quantum entanglement, but also yields
new insights into quantum computation \cite{manko},\cite{Fedele} and quantum
communication.

The paper is organized as follows: In Section \ref{secII}, we will discuss
the analogies between optical transverse modes in a multimode waveguide and
quantum Fock states. In Section \ref{secIII}, the superposition of
transverse modes in a random waveguide is analyzed. In Section \ref{secIV},
the Bell inequality as a criterion of the existence of mode-entangled states
is deduced. In Section \ref{secV}, an analysis of the mode-entangled states
in random waveguides and the correlation properties of group delays is
discussed. Finally, we summarize our conclusions in Section \ref{secVI}.

\section{Analogs of quantum Fock states using optical transverse modes}

\label{secII}

Considering a weakly guiding, symmetric slab waveguide, an optical field in
the propagation direction, longitudinal $z$ direction, is restricted within
the core region, which has the higher refractive index (RI) compared with
that of the cladding. By using the Fock-Leontovich paraxial approximation,
the Maxwell equations for the monochromatic electric field component can be
reduced to the equivalent Schrodinger equation. The reduced field \cite{kov}%
\begin{equation}
\Psi \left( x,y,z\right) =\sqrt{n_{0}}E\left( x,y,z\right) \exp \left(
-ik\int_{0}^{z}n_{0}dz\right)  \label{eq1}
\end{equation}%
satisfies the following equation%
\begin{equation}
\frac{i}{k}\frac{\partial \Psi }{\partial \xi }=-\frac{1}{2k^{2}}\left(
\nabla _{x}^{2}\Psi +\nabla _{y}^{2}\Psi \right) +\frac{1}{2}\left[
n_{0}^{2}-n^{2}\left( x,y,z\right) \right] \Psi =H\Psi  \label{eq2}
\end{equation}%
where $E\left( x,y,z\right) $ is the monochromatic electric field component,
the geometry of the waveguide is defined by the RI profile $n\left(
x,y,z\right) $, $n_{0}=n\left( 0,0,z\right) $, $k=2\pi /\lambda $ (with $%
\lambda $ being the wavelength in free space) and $\xi =\int_{0}^{z}dz/n_{0}$%
. Therefore, when the RI profile $n\left( x,y,z\right) $ is parabolic, the
equation (\ref{eq2}) is similar to the Schrodinger equation for a quantum
harmonic oscillator. Here, instead of Plank's constant $h$, we have the
vacuum light wavelength $\lambda $. And the variable $\xi $ plays the role
of time. Following \cite{marcuse}, coordinate and momentum operators $\hat{X}%
_{i}$ and $\hat{P}_{i}=-\frac{i}{k}\left( \partial /\partial x_{i}\right) $,
($i=1,2$ denote $x$ and $y$ respectively) can be introduced. These operators
obey the standard commutation relations $\left[ \hat{X}_{i},\hat{P}_{j}%
\right] =\left( i/k\right) \delta _{ij}$ and uncertain relations $\left(
\Delta X_{i}\right) ^{2}\left( \Delta P_{i}\right) ^{2}\geq \left(
1/4k^{2}\right) $, where $\left( \Delta X_{i}\right) ^{2}=\left\langle \hat{X%
}_{i}^{2}\right\rangle -\left\langle \hat{X}_{i}\right\rangle ^{2}$ and $%
\left( \Delta P_{i}\right) ^{2}=\left\langle \hat{P}_{i}^{2}\right\rangle
-\left\langle \hat{P}_{i}\right\rangle ^{2}$.

When the optical field propagates in the waveguide that is $z$-invariant, in
other words, the RI profile is uniform along $z$, the propagation can be
equivalently described in the time-independent Schrodinger equation 
\begin{equation}
H\Psi _{n}\left( x,y\right) =\omega _{n}\Psi _{n}\left( x,y\right)
\label{eq3}
\end{equation}%
where $\Psi _{n}\left( x,y\right) $ are a set of eigenmodes corresponding to
a set of discrete eigenvalues $\omega _{n}$. It leads to the expression of
the monochromatic electric field component $E\left( x,y,z\right) $ as
follows 
\begin{equation}
E\left( x,y,z\right) =\sum_{n}C_{n}e^{-i\beta _{n}z}\Psi _{n}\left(
x,y\right)  \label{eq4}
\end{equation}%
where the propagation constants $\beta _{n}$ are given as 
\begin{equation}
\beta _{n}=k\left( n_{0}^{2}-2\omega _{n}\right) ^{1/2}  \label{eq5}
\end{equation}%
As described in \cite{dragoman}, such eigenmodes $\Psi _{n}\left( x,y\right) 
$ are similar to the quantum Fock states. Here we introduce the annihilation
operators $\hat{a}_{i}=\sqrt{k/2}\left( \hat{X}_{i}+i\hat{P}_{i}\right) $
and the creation operators $\hat{a}_{i}^{+}=\sqrt{k/2}\left( \hat{X}_{i}-i%
\hat{P}_{i}\right) $ that obey the boson commutation relations $\left[ 
\hat{a}_{i}^{+},\hat{a}_{j}\right] =\delta _{ij},\left[ \hat{a}_{i},\hat{a}%
_{j}\right] =\left[ \hat{a}_{i}^{+},\hat{a}_{j}^{+}\right] =0$. Application
of the creation and annihilation operators to the Fock states yield 
\begin{eqnarray}
\hat{a}^{+}\left| n\right\rangle &=&\sqrt{n+1}\left| n+1\right\rangle 
\nonumber \\
\hat{a}\left| n\right\rangle &=&\sqrt{n}\left| n-1\right\rangle  \label{eq6}
\\
\hat{a}^{+}\hat{a}\left| n\right\rangle &=&n\left| n\right\rangle  \nonumber
\end{eqnarray}%
where $\left| n\right\rangle $ are the eigenmodes $\Psi _{n}\left( x\right) $%
, $n=0,1,2,\ldots $, (for simplicity only $x$ direction is considered).

It is well known that random perturbations of the geometry of multimode
optical waveguides cause fluctuations of the average arrival time (group
delay) and spread (dispersion) of optical pulse propagating in the
waveguides \cite{rousseau}\cite{jeunhomme}\cite{marcuse2}\cite{arnaud}\cite%
{gloge}. In general, the information given for the description of optical
field propagation in a random waveguide by means of the field $\Psi \left(
x,y,z\right) $ in Eq. (\ref{eq1}) is not complete. The optical field
propagation may be generally described by means of the density matrix
formalism \cite{kov}%
\begin{equation}
\rho =\sum_{mn}\rho _{mn}\left| m\right\rangle \left\langle n\right|
\label{eq7}
\end{equation}%
which satisfies the Liouville equation 
\begin{equation}
i\frac{\partial \rho }{\partial z}=\left[ H,\rho \right]  \label{eq8}
\end{equation}%
The density matrix possesses the usual properties: $\limfunc{Tr}\rho =1,%
\limfunc{Tr}\rho ^{2}\leq 1$ (the equality is true for pure states). The
expectation value of any operator $\hat{Q}$ is given by the trace of the
product of $\rho $ and $\hat{Q}$: $\left\langle Q\right\rangle =\limfunc{Tr}%
\left( \rho \hat{Q}\right) $. Therefore, the utilization of the density
matrix formalism seems to be useful for describing a superposition of modes
in the random waveguide.

\section{Superposition of transverse modes in a random waveguide}

\label{secIII}

In \cite{jianfu}, a full optical method based on the transverse eigenmodes
is proposed to perform the quantum computation, in which $\limfunc{TE}%
\nolimits_{0}$ mode and $\limfunc{TE}\nolimits_{1}$ mode in dual-mode
waveguide are used as qubits to represent logical 0 and 1. In this section,
we will use the density matrix formalism in the analysis of superposition of
these two modes ($\limfunc{TE}\nolimits_{0}$ mode and $\limfunc{TE}%
\nolimits_{1}$ mode) in a random waveguide.

The superposition of the modes can be described as 
\begin{equation}
\Psi \left( x,y,z\right) =C_{0}e^{-i\beta _{0}z}\left| \limfunc{TE}%
\nolimits_{0}\right\rangle +C_{1}e^{-i\beta _{1}z}\left| \limfunc{TE}%
\nolimits_{1}\right\rangle  \label{eq9}
\end{equation}%
where $\beta _{0}$ and $\beta _{1}$ are the propagation constants of the
modes $\left| \limfunc{TE}\nolimits_{0}\right\rangle $ and $\left| \limfunc{%
TE}\nolimits_{1}\right\rangle $, respectively. In the dual-mode waveguide,
the coupling of TE0 mode and TE1 mode is similar to a two-level system. Thus
to describe this kind of coupling, we introduce the Hamiltonian 
\begin{equation}
H=\beta _{0}\left( \hat{a}^{+}\hat{a}+\frac{1}{2}\right) +\beta _{1}\left( 
\hat{b}^{+}\hat{b}+\frac{1}{2}\right) +C_{ab}\hat{a}^{+}\hat{b}+C_{ab}^{\ast
}\hat{b}^{+}\hat{a}  \label{eq10}
\end{equation}%
where $\hat{a}^{+}$ and $\hat{a}$ are the creation and the annihilation
operators of the mode $\left| \limfunc{TE}\nolimits_{0}\right\rangle $, and $%
\hat{b}^{+}$ and $\hat{b}$ are the creation and the annihilation operators
of the mode $\left| \limfunc{TE}\nolimits_{1}\right\rangle $. In the random
waveguide, the random coupling among the guided modes will be caused by the
perturbations in the waveguide geometry. Here we introduce a coupling
coefficient to describe the random coupling. The coupling coefficients are
functions of $z$ coordinate that measures distance along the waveguide axis.
In random waveguides, the coupling coefficient assume the form \cite%
{marcuse3} 
\begin{equation}
C_{ab}=K_{ab}f\left( z\right)  \label{eq11}
\end{equation}%
where $K_{ab}$ is independent of $z$. The function $f(z)$ often describes
the actual shape of the deformed waveguide boundary or the bent waveguide
axis. And it is supposed to be a stationary random variable whose
correlation function is assumed to be Gaussian 
\begin{equation}
\left\langle f\left( z\right) f\left( z-u\right) \right\rangle =\sigma
^{2}e^{-\left( u/D\right) ^{2}}  \label{eq12}
\end{equation}%
where $\left\langle \ldots \right\rangle $ denotes the average over an
ensemble of random realizations, $\sigma $ is the variance and $D$ is the
correlation length of $f(z)$.

To study effects on the superposition of transverse eigenmodes caused by
this kind of randomicity, we introduce the density matrix $\rho $. If we
rewrite the modes $\left| \limfunc{TE}\nolimits_{0}\right\rangle $ and $%
\left| \limfunc{TE}\nolimits_{1}\right\rangle $ as 
\begin{equation}
\left| \limfunc{TE}\nolimits_{0}\right\rangle =\left( 
\begin{array}{c}
1 \\ 
0%
\end{array}%
\right) ,\left| \limfunc{TE}\nolimits_{1}\right\rangle =\left( 
\begin{array}{c}
0 \\ 
1%
\end{array}%
\right) ,  \label{eq13}
\end{equation}%
the density matrix $\rho $ of the mode superposition (\ref{eq9}) can also be
rewritten 
\begin{equation}
\rho =\left( 
\begin{array}{cc}
\left| C_{0}\right| ^{2} & C_{0}C_{1}^{\ast }e^{i\Delta \beta z} \\ 
C_{1}C_{0}^{\ast }e^{-i\Delta \beta z} & \left| C_{1}\right| ^{2}%
\end{array}%
\right)  \label{eq14}
\end{equation}%
where $\Delta \beta =\beta _{1}-\beta _{0}$. The Liouville equation (\ref%
{eq8}) can be written in the equivalent form 
\begin{equation}
i\frac{\partial \rho _{mn}\left( z\right) }{\partial z}=\left( \beta
_{m}-\beta _{n}\right) \rho _{mn}\left( z\right) +\left\langle \left[
V\left( z\right) ,\rho \left( z\right) \right] \right\rangle _{mn}
\label{eq15}
\end{equation}%
where $m,n\in \left\{ 0,1\right\} $, $V\left( z\right) =C_{ab}\hat{a}^{+}%
\hat{b}+C_{ab}^{\ast }\hat{b}^{+}\hat{a}$. By using the method mentioned in %
\cite{marcuse2}, after the optical field propagates a distance of $L$ in the
random waveguide, the density matrix $\rho $ can be described as 
\begin{equation}
\rho =\left( 
\begin{array}{cc}
\left( 1+e^{-2\gamma L}\right) \frac{\left| C_{0}\right| ^{2}}{2}+\left(
1-e^{-2\gamma L}\right) \frac{\left| C_{1}\right| ^{2}}{2} & 
C_{0}C_{1}^{\ast }e^{\left[ i\left( \Delta \beta +\kappa \right) -\gamma %
\right] L} \\ 
C_{1}C_{0}^{\ast }e^{\left[ -i\left( \Delta \beta +\kappa \right) -\gamma %
\right] L} & \left( 1+e^{-2\gamma L}\right) \frac{\left| C_{1}\right| ^{2}}{2%
}+\left( 1-e^{-2\gamma L}\right) \frac{\left| C_{0}\right| ^{2}}{2}%
\end{array}%
\right)  \label{eq16}
\end{equation}%
where 
\begin{eqnarray}
\gamma &=&\sqrt{\pi }\sigma ^{2}De^{-\left( D\Delta \beta /2\right)
^{2}}\left| K_{ab}\right| ^{2}  \label{eq17} \\
\kappa &=&\func{Im}\left[ \sqrt{\pi }\sigma ^{2}De^{-\left( D\Delta \beta
/2\right) ^{2}}\func{erf}\left( iD\Delta \beta /2\right) \left|
K_{ab}\right| ^{2}\right]  \nonumber
\end{eqnarray}%
We have assumed so far that $\Delta \beta \gg \gamma $, $\Delta \beta \gg
\kappa $ and the waveguide is lossless. From Eq. (\ref{eq16}), we can
anticipate if $L\rightarrow \infty $, $\limfunc{Tr}\rho ^{2}\rightarrow 
\frac{1}{2}$, which shows the evolution from the coherence (pure state)
superposition to the incoherence (mixed state) superposition caused by the
perturbation in the random waveguide.

In order to distinguish between the coherence superposition and the
incoherence superposition, we propose a scheme by analyzing the symmetries
of the eigenmodes. In Fig. 1, the profiles of the modes $\left| \limfunc{TE}%
\nolimits_{0}\right\rangle $ and $\left| \limfunc{TE}\nolimits_{1}\right%
\rangle $ are shown, where $\left| \limfunc{TE}\nolimits_{0}\right\rangle $
is symmetric and $\left| \limfunc{TE}\nolimits_{1}\right\rangle $ is
antisymmetric. Y-splitter is a device to split one light beam into two
beams. If the perturbation of Y-splitter's transition is slight (i.e. the
change of propagation constant $\Delta \beta \approx 0$), the splitter can
split the beam with extremely low power loss. When $\left| \limfunc{TE}%
\nolimits_{0}\right\rangle $ and $\left| \limfunc{TE}\nolimits_{1}\right%
\rangle $ are launched in the Y-splitter, $\left| \limfunc{TE}%
\nolimits_{0}\right\rangle $ is split into two symmetric parts, while $%
\left| \limfunc{TE}\nolimits_{1}\right\rangle $ is split into two
antisymmetric parts \cite{tamir}. Therefore, the output states of
Y-splitter's two branches are given explicitly by 
\begin{eqnarray}
\left| +\right\rangle &=&\frac{1}{\sqrt{2}}\left( \left| \limfunc{TE}%
\nolimits_{0}\right\rangle +\left| \limfunc{TE}\nolimits_{1}\right\rangle
\right)  \label{eq18} \\
\left| -\right\rangle &=&\frac{1}{\sqrt{2}}\left( \left| \limfunc{TE}%
\nolimits_{0}\right\rangle -\left| \limfunc{TE}\nolimits_{1}\right\rangle
\right)  \nonumber
\end{eqnarray}%
When the input of Y-splitter is given at a coherence superposition, $\left|
\Psi _{in}\right\rangle =\frac{1}{\sqrt{2}}\left( e^{-i\theta }\left| 
\limfunc{TE}\nolimits_{0}\right\rangle +e^{i\theta }\left| \limfunc{TE}%
\nolimits_{1}\right\rangle \right) $, via expanding the input states in
terms of the output states of the two branches of the splitter, we got $%
\left| \Psi _{in}\right\rangle =\cos \theta \left| +\right\rangle -i\sin
\theta \left| -\right\rangle $, from which the intensities of two branches
can be obtained 
\begin{eqnarray}
\left| \left\langle +|\Psi _{in}\right\rangle \right| ^{2} &=&\cos ^{2}\theta
\label{eq19} \\
\left| \left\langle -|\Psi _{in}\right\rangle \right| ^{2} &=&\sin ^{2}\theta
\nonumber
\end{eqnarray}%
Note that the phase $\theta $ of the input field $\left| \Psi
_{in}\right\rangle $ will cause the variation of intensity in the
Y-splitter's two branches. When the input of Y-splitter is given at an
incoherence superposition, the relationship of the phase between the modes $%
\left| \limfunc{TE}\nolimits_{0}\right\rangle $ and $\left| \limfunc{TE}%
\nolimits_{1}\right\rangle $ is uncertain. The density matrix of the
incoherence superposition is given as 
\begin{equation}
\rho =\sum_{n=0,1}W_{n}\left| \limfunc{TE}\nolimits_{n}\right\rangle
\left\langle \limfunc{TE}\nolimits_{n}\right|  \label{eq20}
\end{equation}%
where $W_{n}=\frac{1}{2}$ are the probabilities for the two modes $\left| 
\limfunc{TE}\nolimits_{0}\right\rangle $ and $\left| \limfunc{TE}%
\nolimits_{1}\right\rangle $. Then the output average intensities in two
branches are $\left\langle +\right| \rho \left| +\right\rangle =\left\langle
-\right| \rho \left| -\right\rangle =\frac{1}{2}$. It shows that when the
input is at the incoherence superposition, no matter how the phase of the
input changes, the output intensities will stay invariable. That is the
essence of our scheme of measuring the intensity difference between
Y-splitter's two branches to distinguish between the coherence superposition
and the incoherence superposition.

Now, we apply the Y-splitter to the analysis of the superposition state that
is the evolution of a coherence superposition state after propagating in a
random waveguide with distance of $L$. We define operators to represent the
operations of a phase controller and Y-splitter 
\begin{eqnarray}
\hat{I}^{+}\left( \theta \right) &=&\hat{P}^{+}\left( \theta \right) \left|
+\right\rangle \left\langle +\right| \hat{P}\left( \theta \right) =\frac{1}{2%
}\left( 
\begin{array}{cc}
1 & e^{2i\theta } \\ 
e^{-2i\theta } & 1%
\end{array}%
\right)  \label{eq21} \\
\hat{I}^{-}\left( \theta \right) &=&\hat{P}^{+}\left( \theta \right) \left|
-\right\rangle \left\langle -\right| \hat{P}\left( \theta \right) =\frac{1}{2%
}\left( 
\begin{array}{cc}
1 & -e^{2i\theta } \\ 
-e^{-2i\theta } & 1%
\end{array}%
\right)  \nonumber
\end{eqnarray}%
where $\hat{P}\left( \theta \right) $ denotes the phase control of the input
state 
\begin{eqnarray}
\hat{P}\left( \theta \right) \left| \Psi \right\rangle &=&\hat{P}\left(
\theta \right) \left( C_{0}\left| \limfunc{TE}\nolimits_{0}\right\rangle
+C_{1}\left| \limfunc{TE}\nolimits_{1}\right\rangle \right)  \label{eq22} \\
&=&C_{0}e^{-i\theta }\left| \limfunc{TE}\nolimits_{0}\right\rangle
+C_{1}e^{i\theta }\left| \limfunc{TE}\nolimits_{1}\right\rangle  \nonumber
\end{eqnarray}%
The control of the phase difference between the two modes $\left| \limfunc{TE%
}\nolimits_{0}\right\rangle $ and $\left| \limfunc{TE}\nolimits_{1}\right%
\rangle $ can be achieved by properly changing the RI of the core layer \cite%
{tamir}.

After the input at a mode superposition state propagates a distance of $L$
in the random waveguide and passes through a phase controller and a
Y-splitter, the intensity difference between the output waveguides can be
obtained by using the density matrix $\rho $ in Eq. (\ref{eq16}) 
\begin{eqnarray}
\left\langle \hat{I}^{+}\left( \theta \right) -\hat{I}^{-}\left( \theta
\right) \right\rangle &=&\limfunc{Tr}\left\{ \rho \left[ \hat{I}^{+}\left(
\theta \right) -\hat{I}^{-}\left( \theta \right) \right] \right\}
\label{eq23} \\
&=&e^{-\gamma L}\left[ C_{0}C_{1}^{\ast }e^{i\left( \Delta \beta +\kappa
\right) L}e^{2i\theta }+C_{1}C_{0}^{\ast }e^{-i\left( \Delta \beta +\kappa
\right) L}e^{-2i\theta }\right]  \nonumber
\end{eqnarray}%
When $C_{0}=C_{1}=\frac{1}{\sqrt{2}}$, $\left\langle \hat{I}^{+}\left(
\theta \right) -\hat{I}^{-}\left( \theta \right) \right\rangle =e^{-\gamma
L}\cos \left[ 2\theta +\left( \Delta \beta +\kappa \right) L\right] $. It is
already obvious that the perturbation in the waveguide geometry may cause
the evolution from the coherence superposition to the incoherence
superposition, namely decoherence, which leads to disappearance of the
intensity difference between two outputs of the Y-splitter.

By using beam propagation method (BPM) \cite{bpm}, we calculate numerically
the behavior of the mode superposition in the dual-waveguide when it is
propagating in the phase controller and the Y-splitter discussed above. The
results are shown in Fig. 2, which show that the intensities of Y-splitter's
two branches vary by changing the RI of the core layer.

\section{Bell Inequality of optical transverse mode entanglement}

\label{secIV}

As shown in \cite{dragoman}, in the Wigner distribution, the optical
transverse modes are similar to quantum Fock states. However, such
similarities are confined to first order coherence (such as
single-particle). The higher order coherence (such as multiparticle) is
regarded as the inherent feature of quantum phenomena, which is nonlocal
quantum correlation shown in the correlation measurement of the quantum
entangled state. The criterion of the existence of the quantum entanglement
is given by the violation of the Bell inequality. In this section, we will
discuss the correlation properties of the optical mode entanglement that is
classical simulation of the quantum entanglement.

We assume that a kind of ``mode-entangled states'' can be generated by means
of some kind of interaction between the optical fields propagating in
multimode waveguides (e.g. the C-NOT gate proposed in \cite{jianfu}). The
mode-entangled states are given as\quad 
\begin{eqnarray}
\left| \Phi _{1}^{\pm }\right\rangle &=&\frac{1}{\sqrt{2}}\left( \left| 
\limfunc{TE}\nolimits_{0}\right\rangle _{c}\left| \limfunc{TE}%
\nolimits_{0}\right\rangle _{t}\pm \left| \limfunc{TE}\nolimits_{1}\right%
\rangle _{c}\left| \limfunc{TE}\nolimits_{1}\right\rangle _{t}\right)
\label{eq24} \\
\left| \Psi _{1}^{\pm }\right\rangle &=&\frac{1}{\sqrt{2}}\left( \left| 
\limfunc{TE}\nolimits_{0}\right\rangle _{c}\left| \limfunc{TE}%
\nolimits_{1}\right\rangle _{t}\pm \left| \limfunc{TE}\nolimits_{1}\right%
\rangle _{c}\left| \limfunc{TE}\nolimits_{0}\right\rangle _{t}\right) 
\nonumber
\end{eqnarray}%
where $c$ and $t$ represent the control and the target fields, respectively.
The states in each waveguide are a mode superposition, but they are
different from a product state 
\begin{equation}
\left| \Psi _{2}\right\rangle =\frac{1}{2}\left( \left| \limfunc{TE}%
\nolimits_{0}\right\rangle _{c}+\left| \limfunc{TE}\nolimits_{1}\right%
\rangle _{c}\right) \left( \left| \limfunc{TE}\nolimits_{0}\right\rangle
_{t}+\left| \limfunc{TE}\nolimits_{1}\right\rangle _{t}\right)  \label{eq25}
\end{equation}%
The difference of $\left| \Phi _{1}^{\pm }\right\rangle $ ($\left| \Psi
_{1}^{\pm }\right\rangle $) and $\left| \Psi _{2}\right\rangle $ can be
obtained not by measuring a single field, but by the correlation measurement
of the control and the target fields. This correlation measurement can show
that the two entangled fields are impartible to some extent. Similar to
quantum entanglement, the violation of the Bell inequality is also used as
the criterion of this impartibility.

To perform the correlation measurement of the Bell inequality, a mode
analyzer with a phase controller is required. The construction mentioned in
the last section consisting of a phase controller and a Y-splitter meets the
requirements. Following Eq. (\ref{eq21}), we define operators $\hat{I}%
_{1}^{\pm }$ and $\hat{I}_{2}^{\pm }$ to represent the mode analyzer's
operations on the control and the target fields, respectively 
\begin{eqnarray}
\hat{I}_{1}^{+}-\hat{I}_{1}^{-} &=&\hat{P}^{+}\left( \theta _{1}\right)
\left( \left| +\right\rangle _{c}\left\langle +\right| _{c}-\left|
-\right\rangle _{c}\left\langle -\right| _{c}\right) \hat{P}\left( \theta
_{1}\right)  \label{eq26} \\
&=&e^{2i\theta _{1}}\left| \limfunc{TE}\nolimits_{0}\right\rangle
_{c}\left\langle \limfunc{TE}\nolimits_{1}\right| _{c}+e^{-2i\theta
_{1}}\left| \limfunc{TE}\nolimits_{1}\right\rangle _{c}\left\langle \limfunc{%
TE}\nolimits_{0}\right| _{c}  \nonumber \\
\hat{I}_{2}^{+}-\hat{I}_{2}^{-} &=&\hat{P}^{+}\left( \theta _{2}\right)
\left( \left| +\right\rangle _{t}\left\langle +\right| _{t}-\left|
-\right\rangle _{t}\left\langle -\right| _{t}\right) \hat{P}\left( \theta
_{2}\right)  \nonumber \\
&=&e^{2i\theta _{2}}\left| \limfunc{TE}\nolimits_{0}\right\rangle
_{t}\left\langle \limfunc{TE}\nolimits_{1}\right| _{t}+e^{-2i\theta
_{2}}\left| \limfunc{TE}\nolimits_{1}\right\rangle _{t}\left\langle \limfunc{%
TE}\nolimits_{0}\right| _{t}  \nonumber
\end{eqnarray}%
where $\hat{P}\left( \theta _{1}\right) $ and $\hat{P}\left( \theta
_{2}\right) $ represent the phase controllers on the control and the target
fields, respectively. As discussed in section \ref{secIII}, when $\theta
_{1} $ and $\theta _{2}$ change, the output intensities of the Y-splitters
will vary correspondingly.

Based on the correlation analysis, we propose an experimental scheme, shown
in Fig. 3, in which the mode-entangled state is generated via the C-NOT gate
proposed in \cite{jianfu}. The input of control field is given at the mode
superposition $\frac{1}{\sqrt{2}}\left( \left| \limfunc{TE}%
\nolimits_{0}\right\rangle +\left| \limfunc{TE}\nolimits_{1}\right\rangle
\right) $ and the input of target field is given at the mode $\left| 
\limfunc{TE}\nolimits_{0}\right\rangle $ or $\left| \limfunc{TE}%
\nolimits_{1}\right\rangle $. Then the output fields of the C-NOT gate are
sent to spatially separated mode analyzers represented by $\hat{I}_{1}^{\pm
} $ and $\hat{I}_{2}^{\pm }$. The detected photocurrents of the mode
analyzers' outputs are passively subtracted and monitored on a spectrum
analyzer (SA) to check for correlations. Therefore, the correlation function
is given by%
\begin{equation}
E\left( \theta _{1},\theta _{2}\right) =\frac{\left\langle \left( \hat{I}%
_{1}^{+}-\hat{I}_{1}^{-}\right) \left( \hat{I}_{2}^{+}-\hat{I}%
_{2}^{-}\right) \right\rangle }{\left\langle \left( \hat{I}_{1}^{+}+\hat{I}%
_{1}^{-}\right) \left( \hat{I}_{2}^{+}+\hat{I}_{2}^{-}\right) \right\rangle }
\label{eq27}
\end{equation}%
Substituting $\left| \Phi _{1}^{+}\right\rangle $ and $\left| \Psi
_{2}\right\rangle $ into Eq. (\ref{eq27}), respectively, we obtain the
correlation functions of the two states 
\begin{eqnarray}
E_{\Phi _{1}^{+}}\left( \theta _{1},\theta _{2}\right) &=&\frac{\left\langle
\left( \hat{I}_{1}^{+}-\hat{I}_{1}^{-}\right) \left( \hat{I}_{2}^{+}-\hat{I}%
_{2}^{-}\right) \right\rangle }{\left\langle \left( \hat{I}_{1}^{+}+\hat{I}%
_{1}^{-}\right) \left( \hat{I}_{2}^{+}+\hat{I}_{2}^{-}\right) \right\rangle }
\label{eq28} \\
&=&\frac{\left\langle \Phi _{1}^{+}\right| \left( \hat{I}_{1}^{+}-\hat{I}%
_{1}^{-}\right) \left( \hat{I}_{2}^{+}-\hat{I}_{2}^{-}\right) \left| \Phi
_{1}^{+}\right\rangle }{\left\langle \Phi _{1}^{+}\right| \left( \hat{I}%
_{1}^{+}+\hat{I}_{1}^{-}\right) \left( \hat{I}_{2}^{+}+\hat{I}%
_{2}^{-}\right) \left| \Phi _{1}^{+}\right\rangle }  \nonumber \\
&=&\cos \left( 2\theta _{1}+2\theta _{2}\right)  \nonumber
\end{eqnarray}%
\begin{eqnarray}
E_{\Psi _{2}}\left( \theta _{1},\theta _{2}\right) &=&\frac{\left\langle
\left( \hat{I}_{1}^{+}-\hat{I}_{1}^{-}\right) \left( \hat{I}_{2}^{+}-\hat{I}%
_{2}^{-}\right) \right\rangle }{\left\langle \left( \hat{I}_{1}^{+}+\hat{I}%
_{1}^{-}\right) \left( \hat{I}_{2}^{+}+\hat{I}_{2}^{-}\right) \right\rangle }
\label{eq29} \\
&=&\frac{\left\langle \Psi _{2}\right| \left( \hat{I}_{1}^{+}-\hat{I}%
_{1}^{-}\right) \left( \hat{I}_{2}^{+}-\hat{I}_{2}^{-}\right) \left| \Psi
_{2}\right\rangle }{\left\langle \Psi _{2}\right| \left( \hat{I}_{1}^{+}+%
\hat{I}_{1}^{-}\right) \left( \hat{I}_{2}^{+}+\hat{I}_{2}^{-}\right) \left|
\Psi _{2}\right\rangle }  \nonumber \\
&=&\cos \left( 2\theta _{1}\right) \cos \left( 2\theta _{2}\right)  \nonumber
\end{eqnarray}%
Then we substitute the correlation functions above into the Bell inequality 
\begin{equation}
\left| B\right| =\left| E\left( \theta _{1},\theta _{2}\right) -E\left(
\theta _{1},\theta _{2}^{\prime }\right) +E\left( \theta _{1}^{\prime
},\theta _{2}^{\prime }\right) +E\left( \theta _{1}^{\prime },\theta
_{2}\right) \right| \leq 2  \label{eq30}
\end{equation}%
This particular Bell inequality is known as Clause-Horne-Shimony-Holt (CHSH)
inequality \cite{bell}. For the entangled state $\left| \Phi
_{1}^{+}\right\rangle $, when we choose $\theta _{1}=\pi /8$, $\theta
_{1}^{\prime }=-\pi /8$, $\theta _{2}=0$, $\theta _{2}^{\prime }=\pi /4$, we
get $\left| B\right| =2\sqrt{2}$. Obviously, by proper choice of the phases $%
\theta _{1}$ and $\theta _{2}$ in Eq. (\ref{eq26}), the correlation of the
analyzers can exhibit a maximum violation of the Bell inequality $\left|
B\right| >2$. However the violation never occurs for the product state $%
\left| \Psi _{2}\right\rangle $.

We have simulated numerically the scheme shown in Fig. 3 by using BPM. The
result is illustrated in Fig. 4. Due to the limitation of the simulation, we
can't get the correlation of the control and the target fields. Therefore,
the experiment is necessary to validate whether there are mode-entangled
states similar to quantum entangled states.

\section{Optical mode-entangled states in random waveguides}

\label{secV}

As shown in section \ref{secIV}, the Bell inequality will be violated in the
correlation measurement of a mode-entangled state. However, we can see from
Eq. (\ref{eq23}) that the violation will vanish due to perturbations of the
random waveguides. The perturbations cause fluctuations of the average
arrival time (group delay) and spread (dispersion) of optical pulse
propagating in the random waveguides. In this section, we will further
discuss the difference of the entangled states $\left| \Phi _{1}^{\pm
}\right\rangle $ ($\left| \Psi _{1}^{\pm }\right\rangle $) and the product
state $\left| \Psi _{2}\right\rangle $ by analyzing the correlation
properties of the group delays. And this difference is another proof of the
existence of the mode-entangled states mentioned in the last section.

After the control and the target fields of the mode-entangled state $\left|
\Phi _{1}^{+}\right\rangle $ propagate respectively in two random waveguides
with the same random characteristic (the variance $\sigma $ and the
correlation length $D$) and the same distance of $L$, the density matrix $%
\rho $ can be described as 
\begin{equation}
\rho _{\Phi _{1}^{+}}=\frac{1}{2}\left( 
\begin{array}{cccc}
1 & 0 & 0 & e^{2\left[ i\left( \Delta \beta +\kappa \right) -\gamma \right]
L} \\ 
0 & 0 & 0 & 0 \\ 
0 & 0 & 0 & 0 \\ 
e^{2\left[ -i\left( \Delta \beta +\kappa \right) -\gamma \right] L} & 0 & 0
& 1%
\end{array}%
\right)  \label{eq31}
\end{equation}%
Due to the perturbations of the random waveguides, the coherent properties
of mode superposition of the control and the target fields will decay
exponentially with increasing the distance of $L$, until the whole state
evolves to an incoherent superposition of $\left| \limfunc{TE}%
\nolimits_{0}\right\rangle _{c}\left| \limfunc{TE}\nolimits_{0}\right\rangle
_{t}$ and $\left| \limfunc{TE}\nolimits_{1}\right\rangle _{c}\left| \limfunc{%
TE}\nolimits_{1}\right\rangle _{t}$. Similarly, the density matrix $\rho $
of the product state $\left| \Psi _{2}\right\rangle $ propagating in the
random waveguides can be described as%
\begin{equation}
\rho _{\Psi _{2}}=\frac{1}{4}\left( 
\begin{array}{cccc}
1 & e^{\left[ i\left( \Delta \beta +\kappa \right) -\gamma \right] L} & e^{
\left[ i\left( \Delta \beta +\kappa \right) -\gamma \right] L} & e^{2\left[
i\left( \Delta \beta +\kappa \right) -\gamma \right] L} \\ 
e^{\left[ -i\left( \Delta \beta +\kappa \right) -\gamma \right] L} & 1 & 
e^{-2\gamma L} & e^{\left[ i\left( \Delta \beta +\kappa \right) -\gamma %
\right] L} \\ 
e^{\left[ -i\left( \Delta \beta +\kappa \right) -\gamma \right] L} & 
e^{-2\gamma L} & 1 & e^{\left[ i\left( \Delta \beta +\kappa \right) -\gamma %
\right] L} \\ 
e^{2\left[ -i\left( \Delta \beta +\kappa \right) -\gamma \right] L} & e^{
\left[ -i\left( \Delta \beta +\kappa \right) -\gamma \right] L} & e^{\left[
-i\left( \Delta \beta +\kappa \right) -\gamma \right] L} & 1%
\end{array}%
\right)  \label{eq32}
\end{equation}%
When $L\rightarrow \infty $, the state $\left| \Psi _{2}\right\rangle $
evolves to an incoherent superposition of $\left| \limfunc{TE}%
\nolimits_{0}\right\rangle _{c}\left| \limfunc{TE}\nolimits_{0}\right\rangle
_{t}$, $\left| \limfunc{TE}\nolimits_{1}\right\rangle _{c}\left| \limfunc{TE}%
\nolimits_{1}\right\rangle _{t}$, $\left| \limfunc{TE}\nolimits_{1}\right%
\rangle _{c}\left| \limfunc{TE}\nolimits_{0}\right\rangle _{t}$ and $\left| 
\limfunc{TE}\nolimits_{0}\right\rangle _{c}\left| \limfunc{TE}%
\nolimits_{1}\right\rangle _{t}$, which is obviously different from the
evolution of the mode-entangled state $\left| \Phi _{1}^{+}\right\rangle $.
And the difference can be shown by a correlation measurement of group delays.

The group delay of an optical pulse in a waveguide can be expressed as 
\begin{equation}
\tau =\frac{L}{c}\frac{d\beta }{dk}  \label{eq33}
\end{equation}%
where $\beta $ is the propagation constant, $c$ is the light velocity and $L$
is the length of waveguide. If we introduce the group delay operator $\hat{%
\tau}$ whose eigenvalues are the group delay $\tau $, the average arrival
time of the optical pulse can be obtained as follows 
\begin{equation}
\left\langle \tau \left( L\right) \right\rangle =\limfunc{Tr}\left( \rho 
\hat{\tau}\right)  \label{eq34}
\end{equation}%
To study the correlation measurement of the group delays, we define the
correlation function between the group delays of the control and the target
fields as 
\begin{eqnarray}
\left\langle \tau _{c},\tau _{t}\right\rangle &=&\left\langle \left( \hat{%
\tau}_{c}-\left\langle \hat{\tau}_{c}\right\rangle \right) \left( \hat{\tau}%
_{t}-\left\langle \hat{\tau}_{t}\right\rangle \right) \right\rangle
\label{eq35} \\
&=&\left\langle \hat{\tau}_{c}\hat{\tau}_{t}\right\rangle -\left\langle \hat{%
\tau}_{c}\right\rangle \left\langle \hat{\tau}_{t}\right\rangle  \nonumber
\end{eqnarray}%
Substituting Eqs. (\ref{eq31}) and (\ref{eq32}) into Eq.(\ref{eq35}), the
correlation functions of the entangled state $\left| \Phi
_{1}^{+}\right\rangle $ and the product state $\left| \Psi _{2}\right\rangle 
$ are obtained 
\begin{eqnarray}
\left\langle \tau _{c},\tau _{t}\right\rangle _{\Phi _{1}^{+}} &=&\limfunc{Tr%
}\left( \rho _{\Phi _{1}^{+}}\hat{\tau}_{c}\hat{\tau}_{t}\right) -\limfunc{Tr%
}\left( \rho _{c1}\hat{\tau}_{c}\right) \limfunc{Tr}\left( \rho _{t1}\hat{%
\tau}_{t}\right)  \label{eq36} \\
&=&\frac{L^{2}}{4c^{2}}\left( \frac{d\beta _{1}}{dk}-\frac{d\beta _{0}}{dk}%
\right) ^{2}  \nonumber \\
&=&\frac{1}{4}\left[ t_{1}\left( L\right) -t_{0}\left( L\right) \right] ^{2}
\nonumber
\end{eqnarray}%
\begin{eqnarray}
\left\langle \tau _{c},\tau _{t}\right\rangle _{\Psi _{2}} &=&\limfunc{Tr}%
\left( \rho _{\Psi _{2}}\hat{\tau}_{c}\hat{\tau}_{t}\right) -\limfunc{Tr}%
\left( \rho _{c2}\hat{\tau}_{c}\right) \limfunc{Tr}\left( \rho _{t2}\hat{\tau%
}_{t}\right)  \label{eq37} \\
&=&0  \nonumber
\end{eqnarray}%
where the reduced density matrices $\rho _{c1}$, $\rho _{t1}$, $\rho _{c2}$
and $\rho _{t2}$ are the partial traces $\limfunc{Tr}_{t}\left( \rho _{\Phi
_{1}^{+}}\right) $, $\limfunc{Tr}_{c}\left( \rho _{\Phi _{1}^{+}}\right) $, $%
\limfunc{Tr}_{t}\left( \rho _{\Psi _{2}}\right) $ and $\limfunc{Tr}%
_{c}\left( \rho _{\Psi _{2}}\right) $, respectively. Here $t_{0}\left(
L\right) $ and $t_{1}\left( L\right) $ are the propagation time in the
waveguide with distance of $L$ for the modes $\left| \limfunc{TE}%
\nolimits_{0}\right\rangle $ and $\left| \limfunc{TE}\nolimits_{1}\right%
\rangle $ respectively. From Eqs. (\ref{eq36}) and (\ref{eq37}), we can see
the difference of the correlation properties of the two states' group
delays. Considerable attention should be paid to that the correlation of the
entangled state will increase, instead of decrease, when the propagation
distance of $L$ increases. Such effects should be able to be observed by
means of the experimental methods shown in \cite{Rokitski}\cite{lou}\cite%
{White}.

\section{Conclusions}

\label{secVI}

We have demonstrated some properties of the ``mode-entangled states'' as the
classical simulation of the quantum entangled states. These properties can
be regarded as the proofs of the existence of the ``mode-entangled states''.
Then two experimental schemes to demonstrate these properties are suggested.
One is based on the violation of the Bell inequality, the other on the
correlation properties of the optical pulses' group delay in random
waveguides. As far as we know, both of the two schemes can be carried out in
current experimental conditions. We are looking forward to performing
relevant experimental schemes.

\begin{acknowledgement}
Supported by the Natural Science Foundation of Zhejiang Province under Grant
No. 601068.
\end{acknowledgement}

\newpage

\begin{description}
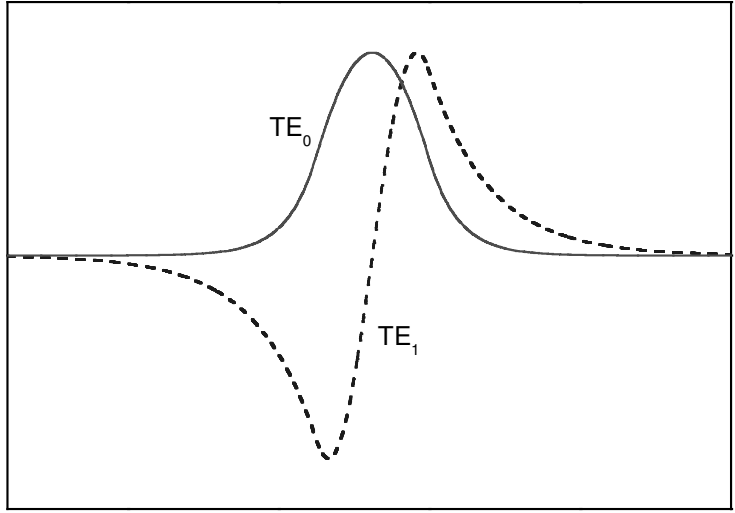
\item Fig. 1: Electric field profiles for the optical transverse modes $%
\limfunc{TE}\nolimits_{0}$ (the symmetric mode) and $\limfunc{TE}%
\nolimits_{1}$ (the antisymmetric mode).

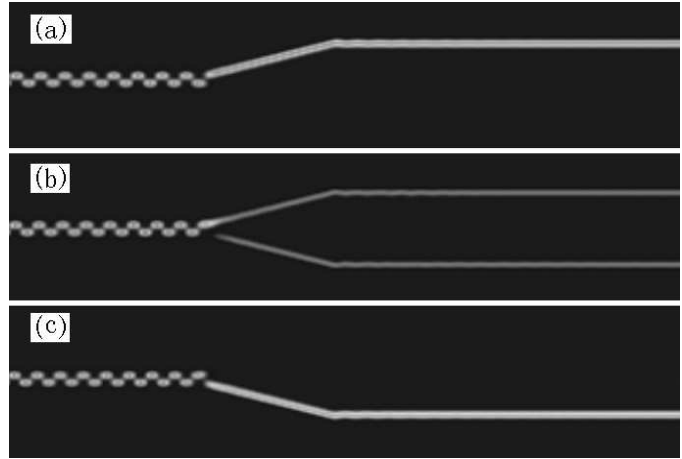
\item Fig. 2: The intensity variances of Y-splitter's two branches by
changing the RI of the core layer (the length $L=1\unit{mm}$): (a) $\Delta
n=0$, (b) $\Delta n=0.0001$, (c) $\Delta n=0.00021$.

\item Fig. 3: Experimental scheme. The input of control field is given at
the mode superposition $\frac{1}{\sqrt{2}}\left( \left| \limfunc{TE}%
\nolimits_{0}\right\rangle +\left| \limfunc{TE}\nolimits_{1}\right\rangle
\right) $ and the input of target field is given at the mode $\left| 
\limfunc{TE}\nolimits_{0}\right\rangle $ or $\left| \limfunc{TE}%
\nolimits_{1}\right\rangle $. Then the output fields of the C-NOT gate are
sent to spatially separated mode analyzers, each of which contains a
Y-splitter and a variable phase control $\theta _{1}$ ($\theta _{2}$). The
detected photocurrents are passively subtracted and monitored on a spectrum
analyzer (SA) to check for correlations.

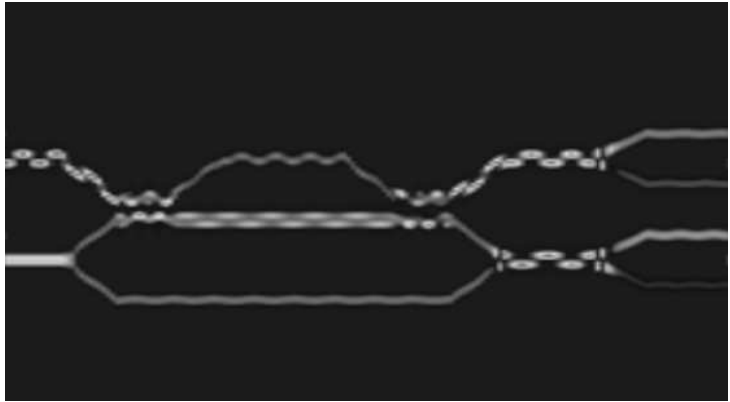
\item Fig. 4: BPM simulation result for the scheme shown in Fig. 3.
\end{description}

\end{document}